\documentclass[usenatbib]{mn2e}
\usepackage{graphicx, amssymb, aas_macros, natbib, bm}
\usepackage{enumitem}

\newcommand{\kms}{{\rm km \, s}^{-1}}

\newcommand{\lcdm}{$\Lambda$CDM}

\newcommand{\lt}{<}
\newcommand{\gt}{>}

\title[Are planes of satellites ubiquitous?]{Are rotating planes of satellite galaxies ubiquitous?}

\author[Phillips et al.]
{John I. Phillips$^1$\thanks{$\!\!$e-mail: johnip@uci.edu}, 
Michael C. Cooper$^1$\thanks{$\!\!$e-mail: cooper@uci.edu}, 
James S. Bullock$^1$, 
\newauthor Michael Boylan-Kolchin$^2$ \\
$\!\!^1$Center for Cosmology, Department of Physics and Astronomy, 
  4129 Reines Hall, University of California Irvine, CA 92697, USA \\
$\!\!^2$Astronomy Department, University of Maryland,
College Park, MD 20742-2421}

\begin{document}

\pagerange{\pageref{firstpage}--\pageref{lastpage}} 
\pubyear{2015}

\maketitle

\label{firstpage}

\begin{abstract} 

  We compare the dynamics of satellite galaxies in the Sloan Digital
  Sky Survey to simple models in order to test the hypothesis that a
  large fraction of satellites co-rotate in coherent planes. We
  confirm the previously-reported excess of co-rotating satellite
  pairs located near diametric opposition with respect to their host, but show that this signal is unlikely to be due to rotating discs
  (or planes) of satellites. In particular, no overabundance of
  co-rotating satellites pairs is observed within $\sim 20^{\circ}-50^{\circ}$ of direct opposition,
  as would be expected for planar distributions inclined relative to
  the line-of-sight. Instead, the excess co-rotation 
  for satellite pairs within $\sim 10^{\circ}$ of opposition is
  consistent with random noise associated with undersampling of an
  underlying isotropic velocity distribution. Based upon the observed
  dynamics of the luminous satellite population, we conclude that at
  most $10\%$ of isolated hosts harbor co-rotating satellite planes (as traced by bright satellites).

\end{abstract}


\section{Introduction}
\label{sec:intro} 

Within the $\Lambda$CDM paradigm, the growth of cosmic structure
proceeds as overdensities collapse into dark matter halos, which
eventually serve as the sites for galaxy formation \citep{white78,
  blumenthal84, davis85}. Over time, the hierarchical accretion and
merging of halos drives the development of substructure, such that
some halos reside within the bounds of larger parent halos
\citep{moore98}. The galaxies hosted by these subhalos are typically
referred to as satellites and are important probes of the evolution of
substructure, as they serve as tracers of dark matter on small scales.

High-resolution $N$-Body and hydrodynamic simulations confirm this
picture of hierarchical structure formation, while also making
predictions regarding the properties of subhalos and the satellite
galaxies they host. In particular, simulated subhalos are not
isotropically distributed with respect to their parent dark matter
halo. Instead, simulations across a broad range of mass scales predict
that satellite galaxies should preferentially lie along orbits aligned
with the major axis of the host halo \citep[e.g.][]{vdb99, knebe04,
  libeskind05, kang07, lovell11, wang13}.
Two possible physical drivers are often associated with this predicted
alignment of substructure with the shape of the larger gravitational
potential: [\emph{i}] preferential destruction (or suppression) of
satellites on orbits anti-aligned with the halo's major axis
\citep{zaritsky99, penarrubia02, pawlowski12b} or [\emph{ii}]
accretion of satellites along preferred directions, perhaps associated
with large-scale filaments \citep{zentner05, libeskind11}.

Observations of galaxies in nearby groups and clusters largely support
the predicted anisotropies found in simulations, such that satellites
in massive dark matter halos are preferentially aligned with the major
axis of the central galaxy and with the larger-scale, filamentary
structure \citep[e.g.][]{west00, plionis03, faltenbacher07, hao11,
  tempel15}.
When pushing to lower-mass, more-isolated halos, studies based on
large spectroscopic samples similarly find that satellites
preferentially reside along the major axis of red (or early-type)
hosts, while the distribution of satellites around blue (or late-type)
hosts is consistent with being isotropic \citep[][but see also
\citealt{zaritsky97}]{brainerd05, sales04, sales09, yang06, azzaro07,
  bailin08}.
This apparent lack of spatial anisotropy for satellites of late-type
hosts is potentially driven by random misalignment between the major
axis of the host's disc and the dark matter halo, such that the
satellites may be aligned with the latter but not the former
\citep{bailin05, libeskind07, deason11}.

In contrast to the satellites of comparable star-forming hosts in the
local Universe, observations of the Local Group suggest that the
spatial distribution of satellites around both the Milky Way and M31
are significantly anisotropic. In particular, the satellites of the
Milky Way preferentially reside near the northern and southern
Galactic poles \citep[i.e.~along the minor axis of the Milky Way
disc,][]{holmberg69}, possibly following a planar arrangement
\citep{lb76, kunkel76, metz07, pawlowski12a}. 
The satellites of M31 are similarly anisotropic in their distribution,
with a large subset belonging to a thin disc or plane
\citep{karachentsev96, koch06, mcconnachie06, metz09, conn13}.

When including velocity information, the anisotropy of the Local Group
satellite distribution becomes even more pronounced, with many of the
Milky Way satellites following polar obits, consistent with a vast,
coherently-rotating plane \citep{metz08, pawlowski13a,
  pawlowski13b}. For M31, a yet more-striking planar structure is
observed, such that a large number of satellites exhibit coherent
rotation along the line-of-sight to the Milky Way, forming a vast
plane with a diameter of $\sim400$~kpc and a thickness of less than
$\sim15$~kpc \citep{ibata13}. While simulations predict that
satellites should preferentially align with the major axis of the host
dark matter halo, the strong anisotropies observed for the Local Group
satellites (especially those around M31) are inconsistent with the
expectations of simulated subhalo populations \citep[][but see also
\citealt{buck15}, who argue that co-rotating planar arrangements of
satellites are predicted by \lcdm]{kroupa05, kroupa10, pawlowski12b,
  pawlowski14c, ibata14b}.

The striking nature of the M31 satellite disc has served as fuel for
many recent studies investigating the possibility of similar,
strongly-anisotropic satellite distributions around galaxies outside
of the Local Group, such as the discovery of possible planar structure
in the satellite distribution of the Centaurus A group \citep{tully15,
  libeskind15}.
In particular, recent analysis of satellite pairs in the Sloan Digital
Sky Survey \citep[SDSS,][]{york00} points towards the possibility of
co-rotating planar satellite structures around nearby massive galaxies
\citep[][hereafter I14]{ibata14}; for $20$ out of $22$ systems, with
satellite pairs located on diametrically-opposed sides of the host
galaxy, I14 detect co-rotation along the line-of-sight, suggesting
that thin satellite planes -- similar to that of M31 -- may be
relatively common. 
Specifically, this result, which is supported by an analysis of the spatial positions of 
photometrically-selected satellite samples, indicates that
$\gtrsim50\%$ of the satellite population may reside in thin
co-rotating planes \citep[][although \citealt{cautun15} argue that the evidence for the ubiquity of such planar structures is not robust]{ibata14c}. 
Given the scarcity of such structures in modern simulations
\citep[][see \citealt{cautun15b} for an argument that the diversity of properties of these structures accounts for their perceived rarity.]{ibata14b, pawlowski14a}, the analysis of I14 poses a strong
test of the $\Lambda$CDM cosmology and thereby warrants further
investigation.

In this paper, we re-examine the kinematic evidence for the existence
of co-rotating planes of satellites around nearby massive hosts by
comparing the coherence of line-of-sight velocities of observed
satellite galaxies to simple models of satellite spatial distributions
and kinematics. The structure of the paper is as follows: in
\S\ref{sec:data}, we discuss the selection of the observational sample
and the measured abundance of co-rotating satellite pairs. In
\S\ref{sec:models}, we introduce our numerical models and compare the
mock observations derived from the models to the observational
data. Finally, in \S\ref{sec:discuss}, we discuss our results in the
context of the search for M31-like planes elsewhere in the
Universe. Throughout our analysis, we employ a $\Lambda$ cold dark
matter ($\Lambda$CDM) cosmology with WMAP7+BAO+$H_{0}$ parameters
$\Omega_{\Lambda} = 0.73$, $\Omega_{m} = 0.27$, and $h =0.70$
\citep{komatsu11}, and unless otherwise noted all logarithms are base
$10$. Throughout the paper, we use the terms ``[satellite] disc'' and
``[satellite] plane'' interchangeably, referring to co-rotating planar
satellite configurations.

\begin{figure}
\begin{center}
  \includegraphics[width=0.9\columnwidth]{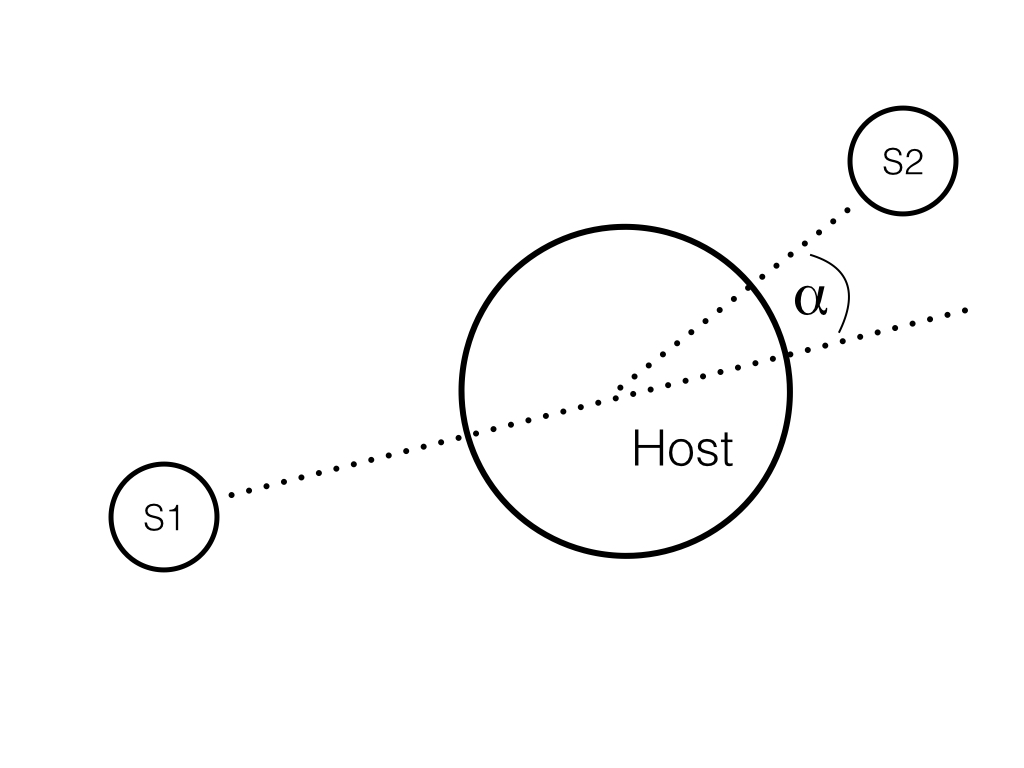}
  \caption{Definition of the opening angle ($\alpha$) between
    satellite pairs (here, S1 and S2), as measured with respect to the
    host galaxy. Each satellite pair has a uniquely defined $\alpha$,
    ranging from $0^{\circ}$ to $180^{\circ}$, such that satellites on
    diametrically-opposed sides of a host correspond to $\alpha =
    0^{\circ}$. }
\label{fig:alpha}
\end{center}
\end{figure}

\begin{figure*}
\begin{center}
  \includegraphics[width=6.5in]{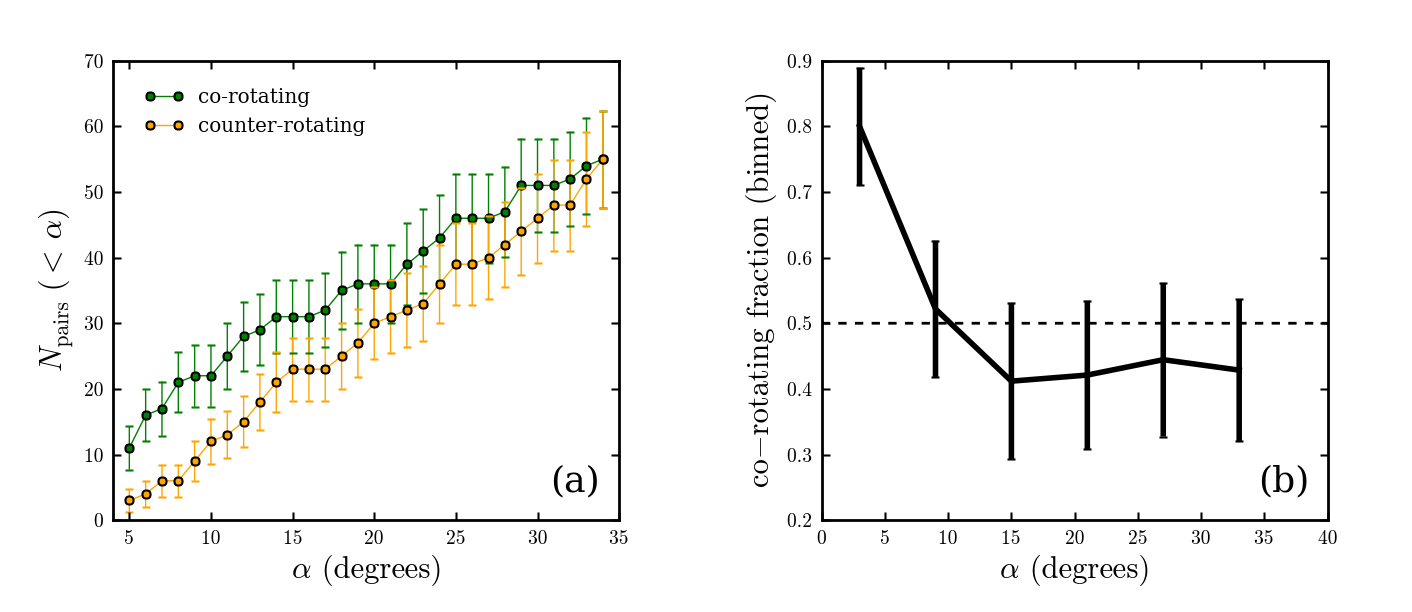}
  \caption{\emph{Left}: the cumulative number of co-rotating (green
    line) and counter-rotating (orange line) satellite pairs as a
    function of opening angle ($\alpha$), including uncertainties
    based on Poisson statistics. \emph{Right}: the fraction of
    co-rotating satellite pairs as a function of opening angle,
    computed in distinct bins of $\alpha$ (bins have width
    $6^{\circ}$), with error bars derived according to the binomial
    theorem. There is a significant excess of co-rotating satellite
    pairs over counter-rotating pairs at small $\alpha$ (i.e.~for
    satellites that are nearly diametrically opposed to each
    other). This overabundance of co-rotating pairs suggests the
    possible presence of coherently-rotating planar satellite
    structures around nearby massive host galaxies (however, see
    Fig.~\ref{fig:full}).}
\label{fig:zoom}
\end{center}
\end{figure*}


\section{Observational data}
\label{sec:data}
\subsection{Sample Selection}

We draw our observational data from Data Release 7
\citep[DR7,][]{abazajian07} of the SDSS, making use of the derived
data products from the NYU Value-Added Galaxy Catalog
\citep[VAGC,][]{blanton05} including absolute $r$-band magnitudes
($M_{r}$) that are $K$-corrected to $z = 0.1$ using \textsc{KCORRECT}
\citep{blanton07}. Throughout this work, we restrict our analysis to
regions of the SDSS where the spectroscopic completeness exceeds
$70\%$ (i.e.~\textsc{FGOTMAIN} $\gt$ 0.7). We also reject all galaxies
with line-of-sight velocity errors greater than $25~{\rm km}/{\rm s}$.

In selecting our galaxy sample, we adhere closely to the procedure of
I14. We select a sample of hosts in the magnitude range $-23 \lt M_{r}
\lt -20$ and within a redshift range of $0.002 \lt z \lt 0.05$. A host
is considered isolated if there are no brighter objects within
$500$~kpc (in projection) on the sky and within $1500~{\rm km}/{\rm
  s}$ in velocity space. Only isolated systems are retained, reducing
the number of hosts to $22,780$ isolated galaxies. From this set of
host systems, we identify galaxies as satellites of a given host if

\vspace*{0.1in}

\noindent (i) their magnitudes fall in the range \\
\indent \indent \indent $M_{r,{\rm host}} + 1 < M_{r,{\rm sat}} < -16$,

\vspace*{0.1in}

\noindent (ii) they are located between $20$~kpc and $150$~kpc from
their \\
\indent host in projected distance ($d_{\rm proj}$), and

\vspace*{0.1in}

\noindent (iii) their velocity offset from the host lies in the
range
$$25~{\rm km}/{\rm s} < |V_{\rm sat} - V_{\rm host}| < 300~{\rm
  km}/{\rm s} \times e^{-(d_{\rm proj}/300~{\rm kpc})^{0.8}}.$$ 

\noindent This velocity bound is taken from I14, and is designed to
reduce the contamination from interlopers in the satellite
sample. Since our interest is in pairs of satellites, we retain only
hosts with two or more satellites. Our final sample contains $427$
such hosts, with $965$ associated satellites. Note that individual
hosts are allowed to harbor more than two satellites; on average, the
SDSS hosts (as well as our model hosts, see \S\ref{sec:models}) have
$2.3$ satellites.

\subsection{Co-rotation signal}
In this subsection, we investigate pairs of satellites for evidence of
co-rotation with respect to their host. To facilitate this, we
introduce the parameter $\alpha$, defined as the angle between the
line extending from one satellite through the host and the position
vector of the second satellite relative to the host, as projected on
the sky (see Figure~\ref{fig:alpha}). We define this opening angle
$\alpha$ such that a satellite pair located on diametrically-opposed
sides of a host will have an opening angle of $0^{\circ}$.
For the duration of this work, we will refer to a satellite pair as
``co-rotating'' if the satellites have opposite-signed (i.e.~one + and
one -) line-of-sight velocity offsets relative to their host and their
associated opening angle ($\alpha$) is less than $90^{\circ}$, or if
they have same-signed velocity offsets relative to their host and
their associated $\alpha$ is greater than $90^{\circ}$. Otherwise, the
satellite pair is deemed to be counter-rotating.

\begin{figure*}
\begin{center}
\includegraphics[width=6.5in]{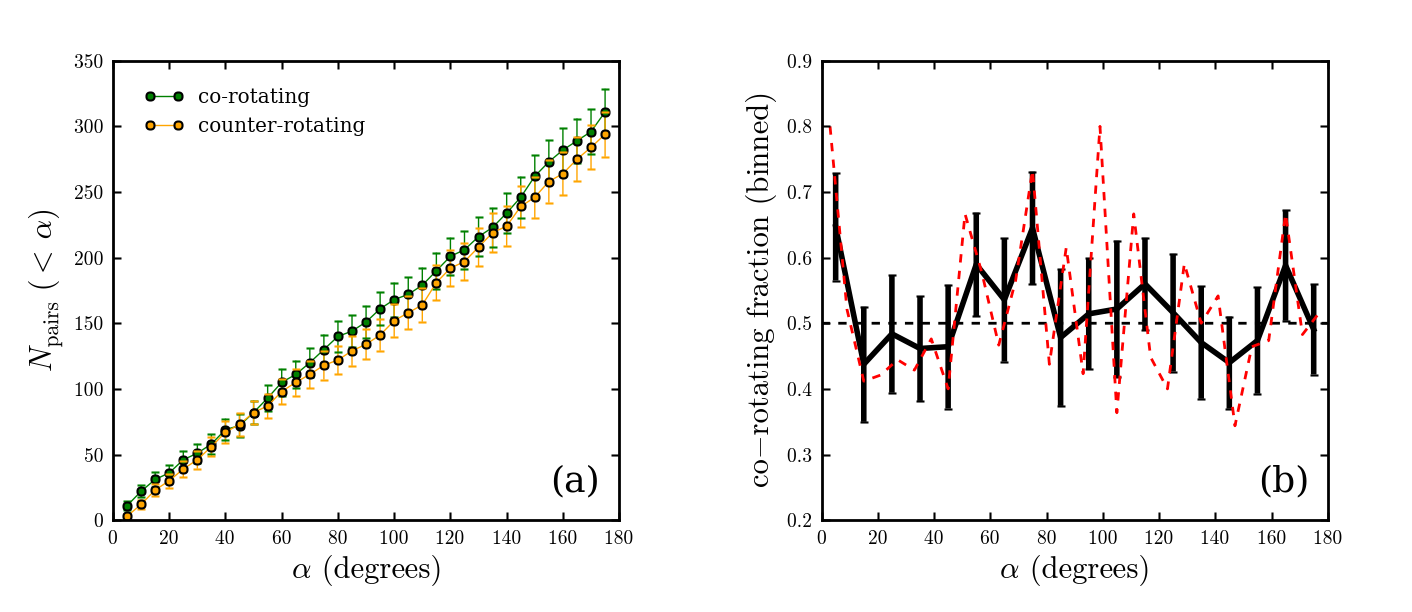}
\caption{\emph{Left}: the cumulative number of co-rotating (green
  line) and counter-rotating (orange line) satellite pairs as a
  function of $\alpha$, including Poisson errors, over the full
  $180^{\circ}$ domain. \emph{Right}: the fraction of co-rotating
  satellite pairs as a function of opening angle, computed in distinct
  bins of $\alpha$ (bins have width $10^{\circ}$), with error bars
  derived according to the binomial theorem (solid black line). To
  facilitate visualization, a coarser binning is adopted than that
  employed in Fig.~\ref{fig:zoom}b; the dash dashed line, however,
  shows the corresponding dependence of co-rotating fraction on
  $\alpha$ utilizing this narrower binning. Over the full range of
  opening angles, the observations are largely consistent with no
  excess of co-rotating satellite pairs, suggesting an isotropic
  velocity distribution.}
\label{fig:full}
\end{center}
\end{figure*}

Figure~\ref{fig:zoom}a shows the cumulative number of co-rotating and
counter-rotating satellite pairs in our sample as a function of
opening angle at $\alpha<35^{\circ}$. At small opening angles there is
a clear excess of co-rotating pairs, as first reported by I14.
The overabundance of co-rotating pairs as a function of $\alpha$ is
better illustrated in Figure~\ref{fig:zoom}b, which shows the fraction
of satellite pairs that are co-rotating as a function of opening
angle, computed in distinct bins of $\alpha$. 
While the surplus of co-rotating pairs at small opening angle is
readily apparent, at $10^{\circ}<\alpha<35^{\circ}$ the signal is
consistent with the sample being divided equally between co-rotating
and counter-rotating (i.e.~a co-rotation fraction of $0.5$, as would
be expected in the absence of any co-rotating structure). At first
glance, the data would seem to indicate the presence of coherently
rotating structures that can only be detected at small values of
$\alpha$ (i.e.~co-rotating planes of satellites viewed close to
edge-on).

When examining the co-rotating fraction of satellite pairs over the
full range of opening angles (i.e.~$0^{\circ}<\alpha<180^{\circ}$),
however, the evidence for planes of satellites is far less
convincing. In Figure~\ref{fig:full}, we show (a) the cumulative
counts of co-rotating and counter-rotating satellite pairs and (b) the
co-rotating fraction, again in discrete bins of $\alpha$, over the
full range of opening angles.\footnote{Note that the co-rotating
  fraction as a function of opening angle is computed using different
  binning procedures in Fig.~\ref{fig:zoom}b and Fig.~\ref{fig:full}b,
  so as to aid in visualization. Our results do not strongly depend on
  how the data are binned. The increase in co-rotating fraction at
  small opening angle appears less significant in Fig.~\ref{fig:full}b
  as a result of allowing satellite pairs with opening angles of $\sim
  10^{\circ}$ in the innermost bin.}
In this light, the excess of co-rotating pairs at $\alpha<10^{\circ}$
seems less likely to result from structured, coherent rotation
associated with planar distributions of satellites. For example,
repeatedly resampling $400$ satellite pairs placed randomly in phase
space (as in our ``isotropic'' model in \S\ref{sec:models}) will
frequently produce satellite samples with excess co-rotating fractions
at random opening angles that are, by definition, not indicative of
any underlying physics. Thus, care must be taken not to overinterpret
the observed overadundance of co-rotating pairs at small angles, if
indeed it is merely the result of random fluctuations associated with
undersampling of an underlying isotropic distribution. The remainder
of this paper will examine the argument that the excess of co-rotating
pairs at small $\alpha$ is significant, and indicative of ubiquitous
coherent co-rotation (similar to that observed for M31's satellite
population) by comparing the SDSS data to statistical models of
satellite kinematics.


\section{Comparison to Toy Models}
\label{sec:models}

In order to gain insight as to whether the data presented in Figures
\ref{fig:zoom} and \ref{fig:full} does indeed argue for the existence
of coherently rotating satellite structures, we compare the SDSS data
to mock observations of simple, idealized ``toy" models of satellite
systems. These models are not intended to give a detailed description
of satellite phase-space distributions.\footnote{In particular, the toy models do not account for potential velocity correlations due to group infall, which \cite{cautun15} argues may be important, nor do they capture the complexities of observing against a background of interloper galaxies with potentially corellated velocities.} However, they do provide a
meaningful basis for comparison to the observational data. We begin by
detailing how each toy model is constructed:

\vspace*{0.1in}

\noindent (i) Isotropic model --- For this model, each host is
randomly assigned $2-5$ satellite galaxies.\footnote{In each case,
  where the model is permitted to have more than two satellites, we
  set the probability of a host having $n$ satellites to be four times
  greater than the probability of having $n+1$ satellites.} The
position and velocity of each satellite, with respect to the host, is
randomly chosen to be between $0$ and $200$~kpc from the host (with
random angular coordinates) and $0$ and $200~\kms$, respectively. The
model is randomly rotated and observed along the $z$ direction ---
i.e.~the $z$ direction is taken to be the line-of-sight and the
$xy$ plane is taken to be the plane of the sky.

\vspace*{0.1in}

\noindent (ii) Disc model --- In this model, $2-5$ satellites are
placed randomly between $0$ and $200$~kpc from the origin on the $xy$
plane (prior to any rotation taking place) and then randomly given a
$z$ coordinate between $-10$ and $10$~kpc. All satellites are assigned
a 3D velocity of $100~\kms$, such that each satellite is in circular
motion about the host, initially rotating in the $xy$ plane. The model is then randomly rotated and viewed
along the $z$ axis. Finally, to mimic observational error in the
line-of-sight velocities, we add to the $z$ component of each
satellite's velocity a random offset drawn from a normal distribution
with a standard deviation of $\sigma_{V} = 20~\kms$.
Our qualitative and quantitative results are not strongly dependent
upon the assumed velocity structure or thickness of the model
satellite discs; since we are only concerned with the sign (+ or -) of
the $z$ component (post-rotation) of the satellite's velocity vector,
the magnitude of that vector -- and any radial dependencies it might
have -- is largely unimportant.

\vspace*{0.1in}

\noindent (iii) M31 model --- This model is based on the position and
velocities of the 13 satellites belonging to the co-rotating plane
identified around M31 by \citet{ibata13}. The three-dimensional
positions of the satellites are taken from \citet{mcconnachie12} and
the line-of-sight velocities are compiled from \citet{mcconnachie12}
and \citet{collins13}. Note that we only consider the 13 satellites
exhibiting coherent rotation; the two satellites aligned with the
planar structure, but with counter-aligned line-of-sight velocities,
are excluded. We assign each mock satellite a velocity, such that the
radial component of its velocity is consistent with the observed
line-of-sight velocities of the true M31 satellites, such that each
satellite's total velocity puts it in circular motion around the
host.\footnote{The origin prior to rotation is taken to be the
  position of a Milky Way observer.} We then randomly select $2-5$
satellites to mock observe (independent of the luminosity of the true
M31 satellites), and the system is randomly rotated and viewed along
the $z$ axis. We again add to the $z$ component of the velocity a
random offset drawn from a normal distribution with
$\sigma_{V}=30~\kms$, representative of measurement error in the
line-of-sight velocity. While constructed using the positions and
velocities of the co-rotating satellites in the M31 plane, it is
useful to note that our model is not a true analog of the observed
system as every member of the M31 planar structure is fainter (by
$\sim1-2$ magnitudes or more) than our satellite luminosity limit
($M_{r} < -16$).

\vspace*{0.1in}

\noindent (iv) Dumbbell model --- In this model, each host is
restricted to exactly two satellites. Once the first satellite is
randomly placed on the $xy$ plane (again, prior to any rotation), the
placement of the second satellite is restricted, such that the opening
angle between the two satellites is less than $10^{\circ}$ when the
system is viewed along the $z$ axis. From there, each satellite is
assigned a $z$ coordinate between $-10$ and $10$~kpc and the system is
subject to random rotation, assigned of line-of-sight velocity errors,
and finally viewed along the $z$ axis. In essence, this model requires
two satellites to be on opposite sides of their hosts and orbiting in
rigid-body rotation. As was the case for the disc model, our results
do not strongly depend on the adopted thickness of the dumbbell.

\vspace*{0.1in}

For each realization of a model, we rotate the system to a random
orientation before observing along the $z$ axis. In our analysis, we
simulate a variety of statistical samples, each consisting of $N=10^6$
model realizations, where most samples include a mix of realizations
drawn from the isotropic model along with one of the other three
models. For example, the $50\%$ disc + $50\%$ isotropic model consists
of $5 \times 10^{5}$ realizations of the disc model and $5 \times
10^5$ realizations of the isotropic model. Note that such a sample
does \emph{not} consist of $10^6$ hosts whose satellites have a $50\%$
probability to be placed in a disc and a $50\%$ probability to be
placed randomly. While we do not explicitly explore cases of this
type, they are equivalent to the cases we explore that have a
percentage disc composition of approximately $p_{\rm sat}^{2}$, where $p_{\rm
  sat}$ is the probability of an individual satellite being placed in a disc. Note that this is only an approximation, as some hosts in our models have three or more satellites.
For example, a case where satellites of each host independently
have a $50\%$ chance of being placed in a disc, would be equivalent to
our $25\%$ disc + $75\%$ isotropic model.

\begin{table}
\centering
\begin{tabular}{l c c c c}
\hline \hline
Model                &$p_{\rm sat}$&$\chi ^2$ & $\tilde{\chi}^2$ & $p$\\ \hline
100\% Disc            				&1.0&460.07 	& 28.76 	&  $\lt 0.001$ 	\\ 
50\% Disc + 50\% Isotropic 		&0.71&179.77 	& 11.39 	&  $\lt 0.001$  	\\
25\% Disc + 75\% Isotropic 		&0.50&71.20 	& 4.45 		&  $\lt 0.001$  	\\
10\% Disc + 90\% Isotropic 		&0.32&22.44  	& 1.40 		& 	0.13	\\ \hline
50\% M31 + 50\% Isotropic  		&0.71&221.79 	& 13.05	&  $\lt 0.001$	\\ \hline
50\% Dumbbell + 50\% Isotropic  &0.71&8.70   & 0.58 		& 	0.89	\\
10\% Dumbbell + 90\% Isotropic  &0.32&21.62    & 1.44 		&	0.12	\\ \hline
100\% Isotropic      			&0&11.40   	& 0.67 &	0.83 \\
\hline \hline
\end{tabular}
\label{tab:chis}
\caption{$\chi ^2$, reduced $\chi ^2$, and $p$ values based on a comparison
  of the observed fraction of co-rotating satellite pairs in the SDSS
  versus that for various statistical samples of model satellite
  distributions as described in \S\ref{sec:models}. Models in which a large
  fraction of hosts harbor discs of satellites (including the model
  based on the M31 plane) are disfavored relative to our dumbbell
  model or the simple isotropic case. Calculations are made taking
  satellite pairs over the full range of $\alpha$ (i.e.~$0^{\circ} \lt
  \alpha \lt 180^{\circ}$), but the $p$-values are largely unchanged
  when we restrict the comparison to a narrower range of opening
  angles (e.g.~$\alpha \lt 35^{\circ}$). Also shown is the probability that an \textit{individual} satellite is found in a disc traced by bright satellites, $p_{\rm sat}$.}   
\end{table}

In Table 1, we present a summary of our analysis of the
$\chi^2$ values describing the goodness of fit for several statistical
samples of modeled systems in comparison to the observed fraction of
co-rotating satellite pairs as a function of opening angle in the SDSS
(see Fig.~\ref{fig:full}b). The number of degrees of freedom that
enter into each $p$-value calculation is based on how restrictive the
model under consideration is: disc models are taken to have one fewer
degree of freedom than the purely isotropic model or the model that is
based on the observed positions of M31 satellites, since satellites
are restricted to being placed in the disc. The dumbbell model
essentially introduces an additional constraint on the disc model, so
dumbbell models have yet one fewer degree of freedom.

\subsection{Disc Model and M31 model}
\label{sec:disc_models}

\begin{figure*}
\begin{center}
\includegraphics[width=4.25in]{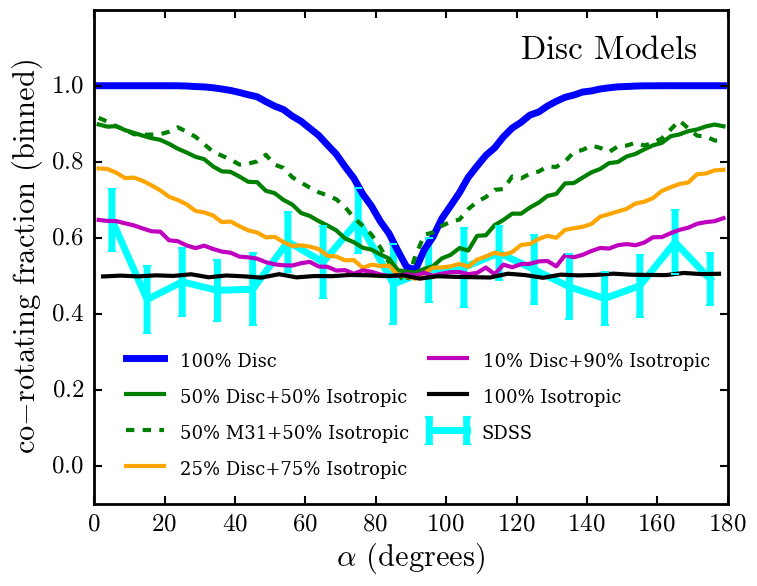}
\caption{The fraction of co-rotating satellite pairs as a function of
  opening angle ($\alpha$) for various versions of the disc model,
  including the model based on the M31 plane, in comparison to the
  observed co-rotating fraction measured in the SDSS (cyan line). The
  observed kinematics of bright satellites in the SDSS are consistent
  with at most $10\%$ of hosts having disc-like or planar satellite
  distributions, with the remainder being distributed isotropically.}
\label{fig:discs}
\end{center}
\end{figure*}

In Figure \ref{fig:discs}, we show the observed fraction of SDSS
satellite pairs that are co-rotating as a function of the opening
angle $\alpha$ in comparison to mock observations of various disc
models. The thick blue line corresponds to a statistical sample comprised
purely of satellite discs, while the green, orange, and magenta lines
correspond to samples composed of $50\%$, $25\%$, and $10\%$ satellite
discs, respectively, with the remainder of each sample consisting of
isotropic satellites. In addition, the dashed green line corresponds
to a sample with satellites for $50\%$ of the simulated hosts
following a M31 model and the other $50\%$ distributed according to an
isotropic satellite population. For comparison, the solid black line
shows the co-rotating fraction for a purely isotropic sample.

We find strong disagreement between the fraction of co-rotating
satellite pairs as a function of opening angle for models in which
$25\%$ or more of hosts harbor planes of satellites in comparison to
that for the observed SDSS sample. In particular, the presence of
inclined planes (relative to the line-of-sight) in the toy models
results in a significant overabundance of co-rotating pairs at
$20^{\circ}\lesssim\alpha\lesssim60^{\circ}$ in comparison to the SDSS
observations.
Overall, the SDSS data agree reasonably well with models where at
most $\sim10\%$ of hosts have satellites residing in planes (or
$\sim90-100\%$ of the hosts have satellites distributed isotropically
in phase space); however, the $100\%$ isotropic model does fail to
reproduce the overabundance of co-rotating pairs at very small opening
angles ($\alpha \lesssim 10^{\circ}$), as measured in the SDSS sample.
As shown in Fig.~\ref{fig:discs}, the M31 sample follows the $50\%$
disc + $50\%$ isotropic sample very closely, as they fundamentally
represent the same satellite arrangement (with the caveat that the
positions and velocities of the satellites in the M31 model are
tailored to match the observed positions of the true M31
satellites). While our results suggest that coherently rotating discs
of bright satellites are not common, the objection could be raised
that the velocity selection criteria used to select the SDSS systems
systematically removes inclined 
satellite planes (i.e.~those systems with satellite pairs at opening
angles of $10^{\circ} \lesssim \alpha \lesssim 170^{\circ}$); we
address this possible selection effect in \S \ref{sec:velocity}.

\subsection{Velocity Modeling and Cuts}
\label{sec:velocity}

As highlighted in \S\ref{sec:disc_models}, those toy models, in which
a high fraction of host galaxies harbor satellite planes, are
disfavored in part due to the lack of excess co-rotating satellite
pairs at intermediate opening angles (i.e.~$20^{\circ} \lesssim \alpha
\lesssim 60^{\circ}$), corresponding to satellites in discs at non-zero
inclination angles.
If our sample selection criteria, in particular our velocity cut, are
biasing us strongly against such systems, we could perhaps reconcile
the apparent discrepancies between the disc models and the SDSS
data. In a simplified test case, where each satellite orbits their
host in a disc at a velocity of $V_0$, imposing a velocity cut of
exactly $V_0$ on the host-satellite velocity offset would retain only
perfectly edge-on discs, leading to a signal much like the one
observed at small opening angles.
Relaxing this velocity cut would permit progressively more face-on
discs, and imposing no velocity cut would in principle permit any disc
inclination angle. In constructing our model, we assigned a
characteristic velocity of $100~\kms$ to the satellites and only
selected those satellites that have a 1D velocity offset (relative to
their host) greater than $\sqrt{2} \times 25~\kms$. Since the toy
model is essentially scale-free, this is equivalent to removing
satellites that have a 1D velocity offset less than $\sqrt{2} \times
25\%$ of the characteristic 3D velocity for satellites of the host ---
i.e.~a velocity threshold of $0.35~V_{0}$ is applied, where $V_0$ is
the characteristic 3D velocity of the satellites. The higher this
velocity cut, the more we would expect a contribution to the
co-rotating fraction at intermediate $\alpha$ from inclined discs to
be suppressed, as such systems would have a significant component of
their velocity moving perpendicular to the line-of-sight.

\begin{figure}
\begin{center}
\includegraphics[width=0.96\columnwidth]{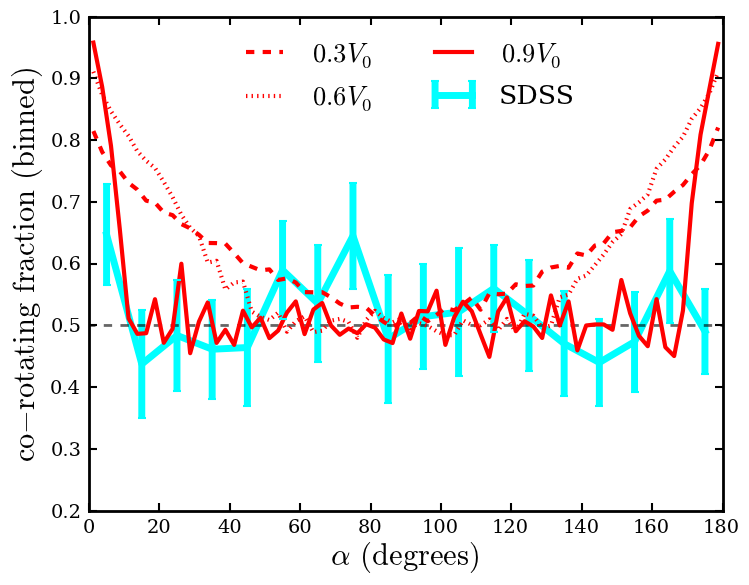}
\caption{The fraction of co-rotating satellite pairs in the SDSS
  satellite sample (cyan line) and in the $25\%$ disc + $75\%$
  isotropic model (red lines), varying the value of the velocity
  selection limit applied to the models. The dotted, dashed, and solid
  red lines correspond to models in which only systems with
  $\Delta V_{los}>0.3$, $0.6$, $0.9~V_{0}$ are included, respectively, and
  where $V_{0}$ denotes the 3D characteristic velocity of satellites
  in the plane/disc. The dramatic overabundance of co-rotating pairs
  at small $\alpha$ in the SDSS data is only captured by models that
  contain nearly exclusively edge-on discs (i.e.~applying the
  $\Delta V_{los}>0.9~V_{0}$ cut).}
\label{fig:cuts}
\end{center}
\end{figure}

In Figure~\ref{fig:cuts}, we demonstrate the impact of altering our
velocity selection criterion on the measured co-rotation fraction
versus opening angle for the $10\%$ disc + $90\%$ isotropic model. The
various red lines range from selection limits of $30\%$ of the 3D
velocity to $90\%$ of the 3D velocity, with the SDSS data included for
comparison (cyan line). Reproducing the lack of excess co-rotation at
$20^{\circ}\lesssim\alpha\lesssim 60^{\circ}$ --- or similarly the
sharpness of the increase in co-rotating satellite pairs at very small
$\alpha$ --- requires a very high velocity cut (i.e.~$\sim 0.9
V_0$). Given that our adopted velocity limit is $\Delta V_{los}
>\sqrt{2} \times 25~\kms$, such a strong selection bias would require
that the characteristic velocity ($V_{0}$) of satellites in planes
would need to be $\sim40~\kms$.
If the planar satellites have such low velocities, a velocity cut of
$\sqrt{2} \times 25~\kms$ would indeed correspond to $0.9$ times the
characteristic velocity of the plane members, and we could confidently
state that we had removed all but edge-on planes.

The ``toy model'' adopted by I14 as a comparison to their measurements
of the co-rotating satellite fraction in the SDSS utilizes exactly
this characteristic velocity ($40~\kms$); as a result, their toy model
only includes edge-on (or nearly edge-one) discs, thereby reproducing
the overabundance of co-rotating satellite pairs at small opening
angles (see their Fig.~1b). 
Assuming a 3D characteristic velocity of $40~\kms$ for satellites in
planes, however, is inconsistent with the broad expectations from
subhalo kinematics in $N$-body simulations as well as the observed
line-of-sight velocities of satellites in the M31 plane, which have a
median value of $|\Delta V_{los}| \sim 92~\kms$.
Moreover, such a low characteristic velocity directly conflicts with
the measured line-of-sight velocity offsets of the co-rotating
satellite pairs identified by I14 (see their Table~1) as well as those
in our sample.

\begin{figure}
\begin{center}
\includegraphics[width=\columnwidth]{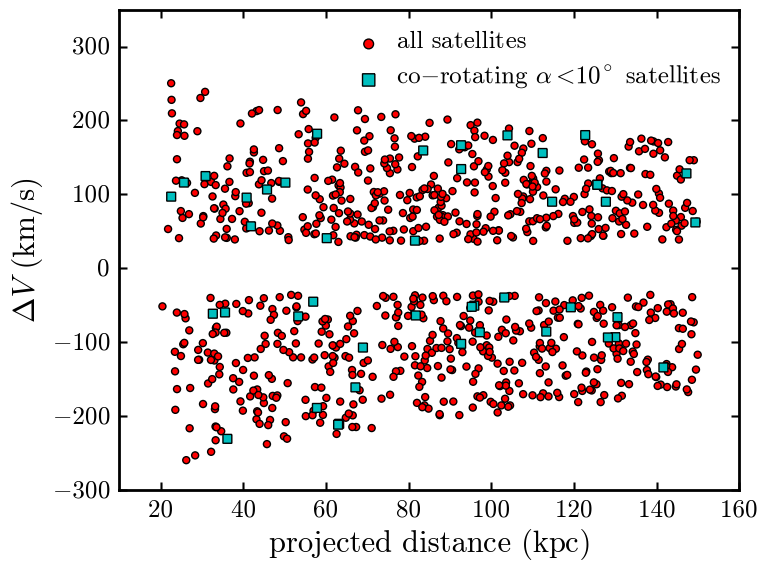}
\caption{The line-of-sight velocity offset for each satellite in the
  SDSS sample relative to its host as a function of its projected
  distance from the host (red circles). Those satellites
  corresponding to co-rotating pairs with opening angles of
  $\alpha<10^{\circ}$ are highlighted as cyan squares.  
  Given that most satellites in the sample (especially those that are
  co-rotating and at small opening angles) have velocity offsets
  greater than the lower selection limit, we conclude that our
  observational sample is not strongly biased against inclined planes
  (or discs).}
\label{fig:dvs}
\end{center}
\end{figure}

Figure~\ref{fig:dvs} shows the host-satellite (line-of-sight) velocity
offset plotted against host-satellite projected distance for each
satellite in our SDSS sample, highlighting those systems belonging to
pairs with $\alpha<10^{\circ}$. The mean absolute value of the 1D
(line-of-sight) velocities for the $44$ satellites in co-rotating
pairs with $\alpha < 10^{\circ}$ is $109.4~\kms$, consistent with that
for the overall sample ($111.4~\kms$). This indicates that a velocity
cut of $\sqrt{2} \times 25~\kms$ would be insufficient to select only
edge-on planes, such that we should also detect co-rotation from
planes at moderate inclination angles (i.e.~yielding an elevated
co-rotating fraction at $10^{\circ} < \alpha < 60^{\circ}$). As a
result, we argue that the observational data are inconsistent with
co-rotating planes being ubiquitous in the local Universe --- at least
with respect to satellites brighter than $M_{r} = -16$.

\subsection{Dumbbell model}

\begin{figure}
\begin{center}
\includegraphics[width=1\columnwidth]{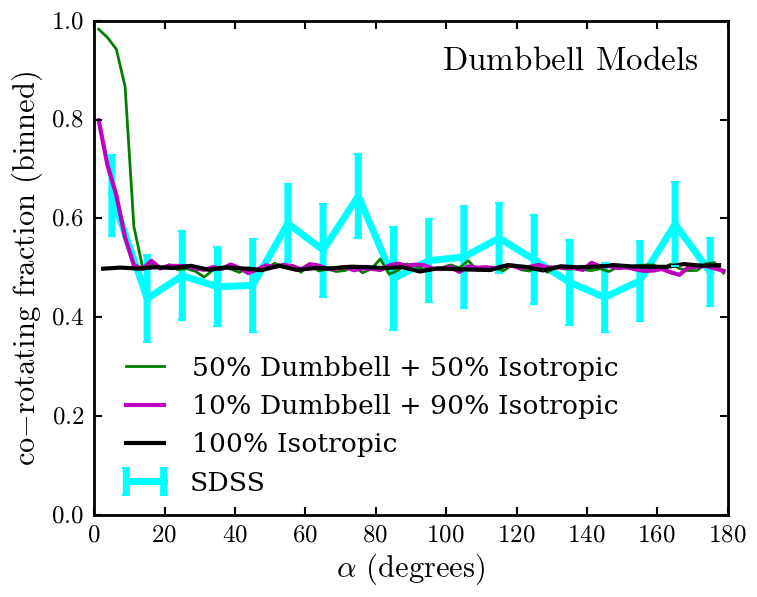}
\caption{The fraction co-rotating satellite pairs as a function of
  opening angle $\alpha$ for the observed SDSS sample (cyan line) and
  various simulated dumbbell samples (green, magenta, and black
  lines). While unlikely to be physically accurate, a model in which a
  small fraction ($\sim10\%$) of massive galaxies host satellites in
  dumbbell-like distributions is able to reproduce the overabundance
  of co-rotating satellites pairs at small opening angles.}
\label{fig:bells}
\end{center}
\end{figure}

Recognizing that the observed variation in co-rotating satellite
fraction with opening angle is strongly inconsistent with discs or
planes of satellites being prevalent, we now discuss a potential
alternative scenario: the dumbbell model (see
\S\ref{sec:models}). Figure~\ref{fig:bells} shows the co-rotating
satellite fraction as a function of opening angle for the SDSS
satellite sample in comparison to two formulations of the dumbbell
model, one with $50\%$ dumbbells (and $50\%$ isotropic satellites) and
one with $10\%$ dumbbells (and $90\%$ isotropic satellites).
Both models are in relatively good agreement with the observational
data. In particular, the dumbbell model is able to reproduce the
observed overabundance of co-rotating pairs at small opening angles
and the corresponding sharp decrease at $\alpha\sim10^{\circ}$.
In \S\ref{sec:discuss}, we address the physical motivation (or lack
thereof) for dumbbell-like satellite configurations. For now, we note
that the dumbbell model yields a better fit to the observed kinematic
data than the disc model. The $p$-value for the toy model with $10\%$
contribution from dumbbells is $0.948$, meaning we fail to reject it
at any confidence level. We also fail, at the $90\%$ confidence level,
to reject the $50\%$ dumbbell model ($p = 0.152$). While we similarly
fail to reject the $10\%$ disc model at $90\%$ confidence ($p=
0.223$), we reject all other disc models at $>99\%$ confidence (see
Table~\ref{tab:chis}).

\section{Discussion}
\label{sec:discuss}

Our analysis is motivated by the work of I14, which found an increase
in the incidence of co-rotation in satellite pairs very near diametric
opposition with respect to their host. If this excess co-rotation is
the result of physical processes, it would tell us something
significant about the co-evolution of satellite systems and the
behavior of dark matter on small scales -- perhaps indicating a
serious problem with the $\Lambda$CDM model. On the other hand, we
must consider the possibility that the co-rotation signal of I14 is
not robust and is the product of random chance. In this section, we
weight these competing possibilities.

In selecting our sample, we adhere closely to the selection criteria
of I14; however, we did not reproduce the I14 satellite sample
exactly. Of the host systems listed in Table~1 of I14, two galaxies
fail our selection criteria. In one of these cases, the host magnitude
fell just outside of our selection window, while the other host had a
neighbor of nearly equal, but slightly brighter magnitude ($\Delta
M_{r} = 0.01$). In addition, our selection includes several systems
not identified by I14, though we do not expect these differences
between our samples to bias our results in any meaningful way. Of the
$22$ satellite pairs with $\alpha<8^{\circ}$ in I14, $20$ ($91\%$)
appear in our sample. Moreover, we reproduce the principal result from
I14, finding an excess of co-rotating satellite pairs located on
diametrically-opposed sides of their host galaxy (i.e.~at small
opening angles). At $\alpha < 8^{\circ}$, we find $21$ out of $27$
satellite pairs to be co-rotating, corresponding to a co-rotating
fraction of $0.78 \pm 0.08\%$.

However, when examining the kinematics of the entire population of
satellite pairs (i.e.~at all opening angles), the dependence of
co-rotating fraction on $\alpha$ is inconsistent with the expectations
from simple models of rotating discs (or planes) of satellites.
For example, at large $\alpha$, we would naively expect an
overabundance of co-rotating satellites mirroring that detected at
small opening angles. \cite{ibata14c} argue that caution must be
taken when considering satellites on the same side of the host, as
relative motion of the satellites with respect to each other (e.g.~in
orbital binary pairs) could dominate the motion of a satellite around
the host, particularly when viewed in projection.
As a result, when examining satellite pairs with large opening angles,
\cite{ibata14c} require that the brighter of the satellite pair be at
least two magnitudes fainter than the host to minimize the impact of
infalling sub-groups; to mitigate deblending problems, identified
pairs closer than $25$~kpc in separation are also excluded. With these
cuts in place, the I14 sample contains $15$ pairs of satellites with
$\alpha>172^{\circ}$, $10$ of which are co-rotating
\citep{ibata14c}. Applying identical cuts to our sample, we find $13$
out of $21$ satellites co-rotating over the same range in $\alpha$,
while removing these additional selection criteria yields $44$ pairs
of satellites, $23$ of which are co-rotating. In each case, the
satellite sample at large $\alpha$ is divided nearly evenly between
co-rotating and counter-rotating pairs, and the co-rotating excess at
large $\alpha$ is significantly less substantive than that at small
$\alpha$. This agrees well with the work of \cite{cautun15}, which
also finds no evidence of excess co-rotation in satellite pairs
located on the same side of their host.

The differences between the expected kinematics of satellites
belonging to discs (or planes) and that observed in the SDSS satellite
sample extend far beyond the large $\alpha$ regime. If the observed
overabundance of co-rotating satellite pairs at small $\alpha$ is
associated with a large population of satellite planes, then we would
expect that planes at non-zero inclination angles to contribute
co-rotating pairs at intermediate opening angles to the observed
sample, resulting in a gradual decrease in the co-rotating fraction at
$0^{\circ} < \alpha < 90^{\circ}$ (see Fig.~\ref{fig:discs}).
Instead, what is observed is a precipitous drop in the co-rotating
fraction, with $21$ out of $27$ ($78 \pm 8.0\%$) satellite pairs found
to be co-rotating at $\alpha < 8^{\circ}$, but only $27$ out of $63$
pairs ($44 \pm 6.2\%$) co-rotating over $8^{\circ} < \alpha <
28^{\circ}$.
Moreover, the dependence of co-rotating fraction on $\alpha$ is quite
noisy (see Fig.~\ref{fig:full}), such that excesses in the co-rotating
fraction -- comparable to that found at $\alpha < 10^{\circ}$ -- are
detected at other relatively narrow ranges of opening angle. For
example, $20$ out of the $31$ satellite pairs with $70^{\circ} <
\alpha < 80^{\circ}$ are co-rotating, which (if significant) is
difficult to reconcile with the predicted kinematics of satellites in
planes. The discrepancies between the data and the disc models are
borne out in the statistical analysis, with models involving more than
$10\%$ contribution from discs (or planes) strongly disfavored (see
Table 1). This is in good agreement with the predictions
of \cite{cautun15}, which argue that only $\sim15\%$ of bright
satellites should share a coherent sense of rotation to within
$25^{\circ}$, as would be required for planar co-rotation. Our
analysis strongly disfavors the claim that planar configurations of
satellites are ubiquitous over the magnitude range ($M_{r}<-16$)
considered.

While the disc model does a relatively poor job of replicating the
observational data, our toy model with satellites arranged in dumbbell
configurations provides a better fit to the observed satellite
kinematics.
By construction, the satellites in the dumbbell model are located at
opposition, such that they able to replicate the sharpness of the
increase in co-rotating fraction at small $\alpha$ while being
consistent with no excess co-rotation at larger opening angles. The
physicality of the dumbbell model is questionable, however, as it is
difficult to imagine a scenario that involves satellites co-rotating
in such a highly-constrained configuration. Additionally, the
satellites belonging to M31's plane are not in a dumbbell
configuration. For these reason, we conclude that the data are not
likely to be the result of dumbbell configurations either.

In the absence of viable alternative models, we argue that the excess
of co-rotating satellite pairs at small $\alpha$ is very likely the
result of random noise. Among the disc models, only the model with a
$10\%$ contribution from satellite discs is a comparably good fit to
the SDSS data as the purely isotropic model. Moreover, 25\% of random
realizations of an isotropic model comprised of $400$ hosts
(i.e.~comparable in number to the observed SDSS sample) yield excess
co- and counter-rotating pairs of satellites comparable in
significance to that measured in the SDSS sample, although just
$\sim1-2\%$ show an overabundance of co-rotating pairs at small
opening angles ($\alpha < 20^{\circ}$).

These results should not be taken to say that there are no planes of
satellites in the Universe. Statistically, the observational data are
roughly consistent with $\lesssim10\%$ of isolated, massive galaxies
playing host to planes, or more specifically planes with multiple
luminous satellites. Our result does not exclude the possibility that
planar structures preferentially populated by faint satellites could
be ubiquitous; if true, this could provide powerful insight into the
formation of planar satellite structures. With the caveat that all of
the satellites under our consideration here are brighter than the
satellites belonging to the M31 plane, we exclude the possibility of
ubiquitous planes analogous to the M31 plane at a very high confidence
level.

\section{Conclusions}

In this work, we investigate the ubiquity of co-rotating planes of
satellites similar to that observed around M31. Using data drawn from
the SDSS, we study the orientation and kinematics of bright ($M_{r} <
-16$) satellite pairs around isolated galaxies, selected to be
roughly analogous to the Milky Way and M31.
By comparing the fraction of co-rotating satellites pairs as a
function of opening angle to the predictions of simple models, we
investigate the signatures of coherent rotation arising from
satellites arranged in planar structures. Our findings are as
follows: 

\begin{itemize}

\item We confirm an excess of co-rotating pairs of satellites at
  opening angles of $\alpha<10^{\circ}$ --- i.e.~satellite pairs
  located on diametrically-opposed sides of their host (see
  Fig.~\ref{fig:zoom}). This overabundance of co-rotating pairs at
  small opening angles, as first identified by I14, has been cited as
  evidence that $50\%$ of satellites belong to coherently-rotating
  planes \citep{ibata14c}. 

\vspace*{0.07in}

\item We find that the excess of co-rotating pairs at small opening
  angles is unlikely to be due to ubiquitous co-rotating planes of
  satellites. While the behavior at small $\alpha$ is suggestive of
  planes aligned along the line-of-sight (i.e.~with an orbital
  inclination of zero), the signal is strongly inconsistent with mock
  observations of satellite galaxy planes (or discs). In particular,
  we find no contribution to the co-rotating fraction from planes
  inclined relative to the line-of-sight (i.e.~with satellites
  configured at opening angles of $10^{\circ} \lesssim \alpha \lesssim
  60^{\circ}$). For our sample of isolated systems, we find that at
  most $\sim10\%$ of hosts harbor satellites planes -- as traced by
  the luminous (LMC-like) satellite population.

\vspace*{0.07in}

\item The excess of co-rotating pairs of satellites at small $\alpha$
  is better fit by a ``dumbbell'' model, where satellites have
  co-rotating partners located opposite the host galaxy (but not a
  true plane). However, this model is very likely unphysical.

\vspace*{0.07in}

\item We do not rule out the possibility that $\sim10\%$ of hosts having co-rotating planes of satellites, or correspondingly $\sim30\%$ of satellites residing in such planes. This case is similar enough to the
  isotropic case that the observed SDSS co-rotating fraction as a
  function of opening angle is roughly consistent with $\lesssim10\%$
  of massive host galaxies harboring planes of satellites.

\end{itemize}

\section{Acknowledgements}

We thank Rodrigo Ibata, Manoj Kaplinghat, and Kev Abazajian for
helpful discussions regarding this work. JSB was supported by NSF
grants AST-1009973 and AST-1009999. Support for this work was provided
by NASA through a Hubble Space Telescope theory grant (program
AR-12836) from the Space Telescope Science Institute (STScI), which is
operated by the Association of Universities for Research in Astronomy
(AURA), Inc., under NASA contract NAS5-26555. This research made use
of Astropy, a community-developed core Python package for Astronomy
\citep{astropy13}.

Funding for the SDSS and SDSS-II has been provided by the Alfred
P. Sloan Foundation, the Participating Institutions, the National
Science Foundation, the U.S. Department of Energy, the National
Aeronautics and Space Administration, the Japanese Monbukagakusho, the
Max Planck Society, and the Higher Education Funding Council for
England. The SDSS Web Site is http://www.sdss.org/.

The SDSS is managed by the Astrophysical Research Consortium for the
Participating Institutions. The Participating Institutions are the
American Museum of Natural History, Astrophysical Institute Potsdam,
University of Basel, University of Cambridge, Case Western Reserve
University, University of Chicago, Drexel University, Fermilab, the
Institute for Advanced Study, the Japan Participation Group, Johns
Hopkins University, the Joint Institute for Nuclear Astrophysics, the
Kavli Institute for Particle Astrophysics and Cosmology, the Korean
Scientist Group, the Chinese Academy of Sciences (LAMOST), Los Alamos
National Laboratory, the Max-Planck-Institute for Astronomy (MPIA),
the Max-Planck-Institute for Astrophysics (MPA), New Mexico State
University, Ohio State University, University of Pittsburgh,
University of Portsmouth, Princeton University, the United States
Naval Observatory, and the University of Washington.


\label{lastpage}
\end{document}